\documentclass[aps,pra,twocolumn,amsfonts,amssymb,amsmath,showpacs,
floatfix,nofootinbib,citesort]{revtex4-2}
\usepackage{mathrsfs}
\usepackage{amsfonts}
\usepackage{amstext}
\usepackage{amsmath}
\usepackage{amssymb}
\usepackage{bm}
\usepackage{CJK}
\usepackage{bbm}
\usepackage[dvips]{graphicx}
\def\qed{\leavevmode\unskip\penalty9999 \hbox{}\nobreak\hfill
	\quad\hbox{\leavevmode  \hbox to.77778em{%
			\hfil\vrule   \vbox to.675em%
			{\hrule width.6em\vfil\hrule}\vrule\hfil}}
	\par\vskip3pt}

\usepackage{amssymb}
\usepackage{graphicx}
\usepackage{graphics}
\usepackage{amsmath}
\usepackage{amsthm}
\usepackage{color}
\usepackage{dsfont}
\usepackage{textcomp}
\usepackage{threeparttable}
\usepackage{booktabs}
\usepackage{diagbox}
\usepackage{float}
\definecolor{darkred}  {rgb}{0.5,0,0}
\definecolor{darkblue} {rgb}{0,0,0.5}
\definecolor{darkgreen}{rgb}{0,0.5,0}
\usepackage{hyperref}
\hypersetup{
	pdftitle = {QRT Proposal},
	pdfauthor = {},
	colorlinks = true,
	urlcolor  = blue,         
	linkcolor = red,     
	citecolor = blue,    
	filecolor = darkred       
}
\usepackage{mathtools}
\def\ra{\rangle}
\def\la{\langle}
\def\bb{\mathbb}
\def\ot{\otimes}
\newtheorem{theorem}{Theorem}

\newtheorem{lemma}{Lemma}

\newtheorem{pro}{Proposition}

\newcommand{\bea}{\begin{eqnarray}}
	\newcommand{\eea}{\end{eqnarray}}
\newcommand{\be}{\begin{equation}}
	\newcommand{\ee}{\end{equation}}
\newcommand{\ba}{\begin{equation}\begin{aligned}}
		\newcommand{\ea}{\end{aligned}\end{equation}}
\newcommand{\rank}{\text{Rank}}

\newcommand{\beax}{\begin{eqnarray*}}
	\newcommand{\eeax}{\end{eqnarray*}}
\newcommand{\bex}{\begin{equation*}}
	\newcommand{\eex}{\end{equation*}}

\theoremstyle{remark}

\newtheorem{example}{Example}

\def\be{\begin{equation}}
	\def\ee{\end{equation}}

\newcommand{\mH}{\mathcal{H}}

\newcommand{\mS}{\mathcal{S}}

\newcommand{\lr}{\rangle\langle}

\newcommand{\tr}{{\rm Tr}}

\newcommand{\mbb}[1]{\mathbb{#1}}

\newcommand{\mbR}{\mathbb{R}}

\begin{document}
	

	\preprint{APS/123-QED}
	\begin{CJK*}{GB}{gbsn}
		\title{Unified entropy entanglement\\}
		

		\author{Wenxue Ren}
		\author{Binghao Li}
		\author{Ruiqun Niu}
		\author{Yu Guo}
		\email{guoyu3@aliyun.com}
		
		\affiliation{School of Mathematical Sciences, Inner Mongolia University, Hohhot, Inner Mongolia 010021, People's Republic of China}
		\affiliation{Inner Mongolia Key Laboratory of Mathematical Modeling and Scientific Computing, Inner Mongolia University, Hohhot, Inner Mongolia 010021, People's Republic of China}
		\author{Shuanping Du}
		\affiliation{School of Mathematical Sciences, Xiamen University, Xiamen, Fujian, 361000, People's Republic of China}


		
		\begin{abstract}
			
			The unified entropy as a generalization of the von Neumann entropy exhibits distinct diversity which contains the Tsallis entropy, the R\'{e}nyi entropy, the von Neumann entropy as special cases. The unified-($r,t$) entropy entanglement with $0<r<1$ and $0< t\leq 1$  was shown to be an entanglement monotone in literature. In this paper, we explore unified-($q,s$) entropy entanglement with $q>1$ and $qs\geq1$ and show that it is also an entanglement monotone and that both of them are monogamous. Going further, we present a global multipartite entanglement measure (GlMEM) based on the unified entropy and it has two subclasses which are classified by the parameters $(q,s)$ and $(r,t)$. Consequently, from the view of the complete multipartite entanglement measure theory, we show that one of them is a complete multipartite entanglement monotone and is not only completely monogamous but also tightly completely monogamous, but the other one is even not complete. We also explore the genuine entanglement measures induced by the unified entropy and the relations with the bipartite entanglement and the global entanglement are discussed, respectively.

		\end{abstract}
		
		
		\maketitle
	\end{CJK*}

	
	\section{Introduction}
	

	Quantum entanglement is not only the key carrier for the fundamental test of quantum mechanics, but also the most important and core physical resource in the theory of quantum information and quantum computing. One of the most difficult yet fundamental issues in the theory of quantum entanglement is quantifying entanglement, also known as entanglement measure\cite{Vedral1997prl,Vedral1998pra,Horodecki2009rmp,Guhne2009pr,Plenio2007qic,Guo2025arxiv}. Entanglement measure is a key indicator for assessing the strength of entanglement correlation and provides an accurate description of the `amount' of entanglement contained in a quantum system. Over the past three decades, people have been constantly seeking and defining the entanglement measures of quantum systems, such as entanglement distillation~\cite{Bennett1996pra}, entanglement cost~\cite{Bennett1996pra,Bennett1996pra2}, entanglement of formation~\cite{Bennett1996pra2,Horodecki2001qic} and concurrence~\cite{Hill1997prl}, R\'{e}nyi entropy entanglement~\cite{Vidal2000}, Tsallis entropy entanglement~\cite{Kim2010pra}, unified-$(r,t)$ entropy entanglement~\cite{Kim2011jpa}, and latent entropy~\cite{Basak2026prl}, etc.

	Among these measures, the ones associated with entropy and its extensions have been widely explored due to the excellent feature of the von Neumann entropy in information theory. The unified entropy~\cite{Hu2006jmp,Rastegin2011jsp} as an extension  of the von Neumann entropy contains, besides the von Neumann entropy, all the well-known types of entropy in quantum information, such as the Tsallis entropy~\cite{Tsallis1988jsp}, the R\'{e}nyi entropy~\cite{Renyi}. Recalling that, the unified-$(r,t)$ entropy is defined by~\cite{Hu2006jmp} (in this paper we also call it unified entropy for short sometimes) 
	\be\label{unified entropy}
	S_{r,t}(\rho):=\frac{1}{(1-r)t}\left[\left(\tr\rho^r\right)^t-1\right],\quad r>0, r\neq 1, t\neq0.
	\ee
	In 2011, Rastegin~\cite{Rastegin2011jsp} showed that $S_{r,t}$ are concave for $0<r<1$ and $t\leq 1$, and that it is subadditive for $r>1$ and $t>1/r$.
	Consequently, Kim and Sanders introduced the unified-$(r,t)$ entanglement in Ref.~\cite{Kim2011jpa}:
	\bea\label{uep}
	E_{r,t}(|\psi\rangle):=S_{r,t}(\rho^A),\quad 0<r<1, 0< t\leq 1,
	\eea
	for pure state and by the convex-roof extension for the mixed states.

	In this paper, we show that as a dual of $E_{r,t}$, the unified-$(q,s)$ entropy with $q>1$ and $qs\geqslant1$ can also define an entanglement monotone which we denote by $E_{q,s}$ (Hereafter, for convenience, the terminology unified-$(r,t)$ entropy refers to the case when $0<r<1$ and $0< t\leq 1$ while the terminology unified-$(q,s)$ entropy refers to the case when $q>1$ and $qs\geq1$). Moreover, both $E_{r,t}$ and $E_{q,s}$ can be extended to quantifying multipartite entanglement. 

	One issue that is closely related to entanglement measure is the monogamy of the entanglement. Monogamy is a fundamental property of quantum entanglement, which describes the distribution law of quantum entanglement in multipartite systems~\cite{Terhal2004}. Specifically, monogamy indicates that if the entanglement between some subsystems in a quantum system is stronger, then the entanglement between these subsystems and other subsystems is weaker. Monogamy provides an important normative standard for entanglement measures. A reasonable entanglement measure should satisfy monogamy; otherwise, it may not accurately reflect the physical characteristics of quantum entanglement. For example, many entanglement measures, such as all the other measures
	that defined by the convex-roof extension~\cite{Guo2019pra}, the squashed
	entanglement~\cite[Theorem 8]{Koashi2004pra} and one-way
	distillable entanglement~\cite[Theorem 6]{Koashi2004pra} are
	monogamous (see~\cite{Guo2025arxiv} for detail).

	For the multipartite entanglement measure (MEM), the theory of the complete MEM and the complete monogamy has been presented in order to characterize the multipartite entanglement in a unified and fine-grained manner~\cite{Guo2020pra,Guo2021qst,Guo2022entropy,Guo2023pra,Guo2024pra,Guo2024rip,Guo2025qip,Guo2025arxiv}. So we also discuss the completeness and the complete monogamy of the two types of GlMEMs we proposed. Consequently, we show that one of them is complete and not only completely monogamous but also tightly completely monogamous in the sense of complete MEM theory, and that the other one is not even complete.

	Throughout this paper, we denote by $A_1A_2\cdots A_n$ an $n$-partite quantum system with the state space $\mH^{A_1A_2\cdots A_n}=\mathcal{H}^{A_1}\otimes
	\mathcal{H}^{A_2}\otimes\cdots\otimes\mathcal{H}^{A_n}$, where $\mH^{A_i}$'s are Hilbert spaces with finite dimension, and by $\mS^{X}$ we denote the set of all density
	operators (or called states) acting on $\mH^{X}$. The superscript or subscript $X$ always denotes the corresponding system. 
	For example, the state in $\mS^{X}$ is denoted by $\rho^X$ (or $\rho_X$ sometimes), and is also denoted by $\rho$ for simplicity whenever the associated system $X$ is clear from the context. $X_1|X_2| \cdots |X_{k}$ denotes the $k$-partition of $A_1A_2\cdots A_n$ (or subsystem of $A_1A_2\cdots A_n$ sometimes), $k\leq  n$. For instance, partition $AB|C|DE$ is a $3$-partition of the 5-particle system $ABCDE$ with $X_1=AB$, $X_2=C$ and $X_3=DE$. The case of $k=n$ is just the original $n$-particle system without any other partition, namely, $A_1A_2\cdots A_n$ means $A_1|A_2|\cdots |A_n$. So, in general $k< n$ unless otherwise specified.
	We denote the set of all the $k$-partitions of $A_1A_2\cdots A_n$ by $\Gamma_k$ as in Ref.~\cite{Guo2024pra}, $2\leq  k<n$, i.e., $\Gamma_k=\{\gamma_i\}$, where $\gamma_i=X_{1(i)}|X_{2(i)}|\cdots|X_{k(i)}$.

	The structure of this article is as follows. Sec. II is a preliminary in which we introduce the monogamy relation of the bipartite entanglement measure, the complete MEM, the coarsening relation of multipartite partition and the complete monogamy relation of MEM, respectively. In Sec. III, we show that the unified-$(q,s)$ entropy is strictly concave, from which we conclude that the bipartite unified-$(q,s)$ entropy entanglement is an entanglement monotone and it is monogamous. A comparison between unified-$(q,s)$ entropy entanglement and unified-$(r,t)$ entropy entanglement is given. Sec. IV is devoted to investigating the GlMEMs in terms the unified entropy. Two ways of quantifying global multipartite entanglement are obtained, and the completeness and complete monogamy relation are discussed accordingly. We then explore the genuine entanglement measure induced by the unified entropy in Sec. V. Finally, Sec. VI summarized the entire paper. For more clarity, all the proofs of our results are given in Appendix.

	
	\section{Preliminary}
	
	
	We review some concepts which are necessary in our discussion. Especially the theory of the complete GlMEM and the complete monogamy relation were established recently although the multipartite entanglement is well-known. In addition, the term reduced function is also a concept that fixed recently. Despite the monogamy relation of entanglement has been explored more than 25 years, the improved definition of the monogamy relation and the monogamy criterion of the convex-roof extended bipartite entanglement monotone may be not well-known. We thus introduce these new concepts and results here for convenience.

	\subsection{Unified entropy}

	Throughout this paper, we denote by 
	\bea
	S_{q,s}(\rho)=\frac{\left(\tr\rho^{q}\right)^s-1}{(1-q)s},\quad q>1, qs\ge 1
	\eea
	the unified-$(q,s)$ entropy and by 
	\bea
	S_{r,t}(\rho)=\frac{\left(\tr\rho^{r}\right)^t-1}{(1-r)t},\quad 0<r<1, 0<t\le 1
	\eea
	the unified-$(r,t)$ entropy. We summarize below the special cases of these two family of unified entropy.

	\begin{enumerate}
		\item[(i)] von Neumann entropy.
		As $q\to1$, the unified $(q,s)$-entropy converges to the von Neumann entropy~\cite{Kim2011jpa}:
		\begin{equation}
			\lim_{q\to1}S_{q,s}(\rho)=-\tr(\rho\log\rho)=S(\rho).
		\end{equation}

		\item[(ii)] Tsallis entropy.
		The unified $(q,s)$-entropy and the unified $(r,t)$-entropy tend to the Tsallis entropy as $s\to1$ and $t\to1$~\cite{Landsberg1998pla}, respectively:
		\beax
		\lim_{s\to1}S_{q,s}(\rho)=\frac{\tr\rho^q-1}{1-q}=T_q(\rho),\\
		\lim_{t\to1}S_{r,t}(\rho)=\frac{\tr\rho^r-1}{1-r}=T_r(\rho).
		\eeax

		\item[(iii)] R\'{e}nyi entropy.
		The unified $(q,s)$-entropy and the unified $(r,t)$-entropy converge to the R\'{e}nyi entropy as $s\to0$ and $t\to0$~\cite{Horodecki1996pla}, respectively:
		\beax
		\lim_{s\to0}S_{q,s}(\rho)=\frac{\log\tr\rho^q}{1-q}=R_q(\rho),\\
		\lim_{t\to0}S_{r,t}(\rho)=\frac{\log\tr\rho^r}{1-r}=R_r(\rho).
		\eeax	
	\end{enumerate}

	\subsection{Monogamy relation of bipartite entanglement measure}

	If the function $E: \mS^{AB}\to\mathbb{R}^{+}$ satisfies the following conditions, it is called a bipartite entanglement measure~\cite{Vedral1997prl}: {(E1)} For all separable states $\sigma^{AB}\in \mS^{AB}$, $E(\sigma^{AB})=0$;
	{(E2)} $E$ does not increase under local operations and classical communication (LOCC), that is, for any LOCC mapping $\varepsilon$, it satisfies
	$E\left[ \varepsilon(\rho^{AB})\right] \leq E(\rho^{AB})$ for all $\rho^{AB}\in \mS^{AB}$.	 
	If $E$ is a convex function and it does not increase on average under LOCC, then $E$ is called an entanglement monotone~\cite{Christandl2004jmp,Eisert2001pra,Horodecki2000prl}.
	Here, $E$ does not increase on average under LOCC means that, if $\rho$ is converted to $\rho_{j}$ with some probability $p_j$ under LOCC, then
	$\sum_{j}p_jE(\rho_{j})\leq  E\left(\rho\right)$~\cite{Plenio2005prl,Vedral1997pra,Vedral1998pra,Vidal2000}.
	
	For any bipartite entanglement measure $E$, the convex-roof extension of $E$, denoted by $E_F$, is defined as
	\bea\label{convex-roof}
	E_F(\rho^{AB}):= \min\limits_{\{p_j,|\psi_j\ra\}}\sum\limits_jp_j E(|\psi_j\ra^{AB}),
	\eea
	where the minimum runs over all the pure states ensembles $\{p_j, |\psi_j\ra\}$ of $\rho^{AB}$. We abbreviate $E_F$ to $E$ hereafter. For any bipartite entanglement measure $E$ on $\mS^{AB}$, if there exists a non-negative function $h: \mS^A\rightarrow\mbR^+$ such that 
	\bea\label{h}
	h\left( \rho^A\right) = E\left( |\psi\lr\psi|^{AB}\right), \quad \rho^A=\tr_B|\psi\lr\psi|^{AB}
	\eea 
	when it is evaluated for the pure states, we call $h$ the {reduced function} of $E$~\cite{Guo2023njp,Guo2024pra,Guo2024rip}. 
	$E_F$ is an entanglement monotone if and only if the reduced function $h$ is concave~\cite{Vidal2000}, i.e.,
	$h[\lambda\rho_1+(1-\lambda)\rho_2]\geq\lambda h(\rho_1)+(1-\lambda)h(\rho_2)$
	for any states $\rho_1$, $\rho_2$, and any $0\leq\lambda\leq1$.

	For a bipartite entanglement measure $E$, $E$ is said to be monogamous if~\cite{Coffman2000pra,Koashi2004pra}
	\bea\label{monogamy1}
	E(\rho^{A|BC})\geq E(\rho^{AB})+E(\rho^{AC})
	\eea
	for any $\rho\in\mS^{ABC}$, where the vertical bar indicates 
	the bipartite split across which we will measure the  
	correlation (Hereafter, when we discuss the entanglement measure of a given state $\rho\in\mS^{A_1A_2\cdots A_n}$, $\rho^X$ always denotes some reduced state of $\rho$ whenever $X$ is some subsystem of $A_1A_2\cdots A_n$). In Ref.~\cite{GG2018q}, we improved the definition of monogamy as: A bipartite measure of entanglement $E$ is monogamous if for any $\rho^{ABC}\in\mS^{ABC}$ that satisfies the {disentangling condition},~i.e.,    
	\bea\label{cond}
	E( \rho^{A|BC}) =E( \rho^{AB}),
	\eea
	we have that $E(\rho^{AC})=0$.
	With respect to this definition, a continuous measure $E$ is monogamous according to this definition if and only if there exists $0<\alpha<\infty$ such that
	\bea\label{power}
	E^\alpha( \rho^{A|BC}) \geq E^\alpha( \rho^{AB}) +E^\alpha( \rho^{AC})	
	\eea
	for all $\rho$ acting on the state space $\mH^{ABC}$ with fixed $\dim\mH^{ABC}=d<\infty$ (see Theorem 1 in Ref.~\cite{GG2018q}).
	It was shown in~\cite{Guo2019pra}, if the reduced function of $E$ is strictly concave, $E_F$ is monogamous according to definition in~\eqref{cond}.

	\subsection{Coarsening relation of multipartite partition}

	Let $\gamma$ and $\gamma'$ be two partitions of $A_1A_2\cdots A_n$ or its subsystems, where $1\le k\leq  n$, $1\le l\leq n$. We denote by~\cite{Guo2022entropy,Guo2024pra,Guo2025arxiv}
	\bea
	\gamma\succ^a \gamma', ~
	\gamma\succ^b \gamma',~
	\gamma\succ^c \gamma'
	\eea 
	if $\gamma'$ can be obtained from $\gamma$
	by the corresponding operation below:
	\begin{itemize}
		\item[(a)] discarding some subsystem(s) of $\gamma$,
		\item[(b)] combining some subsystems of $\gamma$,
		\item[(c)] discarding some subsystem(s) of some subsystem(s) $X_t$ provided that $\gamma=X_1|X_2| \cdots| X_{k}$, $X_{t}=A_{t(1)}A_{t(2)}\cdots A_{t(f(t))}$ with $f(t)\geq2$, $1\leq  t\leq  k$,
	\end{itemize}
	respectively. For example,
	$A|B|C|D\succ^a A|B|D\succ^a B|D$,
	$A|B|C|D\succ^b AC|B|D\succ^b AC|BD$, 
	$A|BC\succ^c A|B$.
	We call $\gamma'$ is coarser than $\gamma$ if 
	$\gamma'$ can be obtained from $\gamma$
	by one or some of the ways in item (a)-item (c), and we denote it by 
	$\gamma\succ \gamma'$
	uniformly.

	Furthermore, if $\gamma\succ \gamma'$,
	we denote by 
	\bea \label{Xi}
	\Xi(\gamma- \gamma')
	\eea
	the set of
	all the partitions that are coarser than $\gamma$ but (i) neither coarser than $\gamma'$ nor the one from which one can derive $\gamma'$ by the coarsening means, and (ii) if it includes some or all subsystems of $\gamma'=Y_1|Y_2| \cdots |Y_{l}$, then all the subsystems $Y_j$'s included are regarded as one subsystem and (iii) if $\gamma'=Y_1|Y_2| \cdots |Y_{l}$ and $\gamma=X_{l}|X_2|\cdots| X_{k}$ with $Y_1|Y_2| \cdots |Y_{l}=X_1|X_2| \cdots|X_{l-1}|X_{l}\cdots X_{k}$, $\Xi(\gamma- \gamma')$ contains only $X_{l}|\cdots| X_{k}$ and the one coarser than it. For example, $\Xi(A|B|C|D-A|BCD)=$$\{B|C|D$, $B|CD$, $BC|D$, $C|BD$, $B|C$,
	$C|D$, $B|D\}$.

	\subsection{Complete multipartite entanglement measure}

	Recall that, a function $E^{(n)}: \mS^{A_1A_2\cdots A_n}\to\mbb{R}^{+}$ is called a $n$-partite entanglement measure~\cite{Horodecki2009rmp,Hongyan2012pra,Guo2020pra} if it satisfies: {(E1)} $E^{(n)}(\rho)=0$ if $\rho$ is fully separable; {(E2)} $E^{(n)}$ cannot increase under $n$-partite LOCC. In addition, $E^{(n)}$ is said to be a $n$-partite entanglement monotone if it is convex and does not increase on average under $n$-partite stochastic LOCC.

	Going further, an MEM $E^{(n)}$ is called a {unified}
	GlMEM if it satisfies the unification condition~\cite{Guo2020pra,Guo2024rip,Guo2025arxiv}:
	(i) (additivity) 
	\bea \label{supadditivity}
	&&E^{(n)}(A_1A_2\cdots A_k\ot{A_{k+1}\cdots A_n})\nonumber\\
	&=& E^{(k)}({A_1A_2\cdots A_k})+E^{(n-k)}({A_{k+1}\cdots A_n}),
	\eea 
	holds for all $\rho^{A_1A_2\cdots A_k}\ot\rho^{A_{k+1}\cdots A_n}\in\mS^{A_1A_2\cdots A_n}$, hereafter $E^{(n)}(X)$ refers to $E^{(n)}(\rho^X)$ and $E^{(1)}=0$;
	(ii) (symmetry) $E^{(n)}({A_1A_2\cdots A_n})=E^{(n)}({A_{\pi(1)}A_{\pi(2)}\cdots A_{\pi(n)}})$,
	for any $\rho\in\mS^{A_1A_2\cdots A_n}$ and any permutation $\pi$;
	(iii) (coarsening monotone) 
	\bea\label{coarsen}
	E^{(k)}(\gamma)\geq E^{(l)}(\gamma')
	\eea
	holds for all $\rho\in\mS^{A_1A_2\cdots A_n}$ whenever $\gamma\succ^a \gamma'$,
	where $\gamma$ and $\gamma'$ are two partitions of $A_1A_2\cdots A_n$ or subsystem of $A_1A_2\cdots A_n$.
	$E^{(n)}$ is called a {complete}
	GlMEM if it satisfies both the unification condition above and the hierarchy condition~\cite{Guo2020pra}: 
	(iv) (tight coarsening monotone) Eq.~\eqref{coarsen} 
	holds for all $\rho\in\mS^{A_1A_2\cdots A_n}$ whenever $\gamma\succ^b \gamma'$.
	By definition, a unified MEM $E^{(n)}$ in fact refers to a family of measures $\{E^{(k)}: 2\leq k\leq n\}$. For example, if $E^{(n)}$ is unified, and we consider the tripartite system, then for any state, $E^{(3)}(ABC)\geq E^{(2)}(AB)$, and if it is complete, then $E^{(3)}(ABC)\geq E^{(2)}(A|BC)$.
	Not all MEMs are unified and some unified GlMEMs are not complete~\cite{Guo2024rip}.

	\subsection{Complete monogamy relation of the multipartite entanglement measure}

	In Ref.~\cite{Guo2020pra}, in order to characterize the distribution of entanglement in a ``complete'' sense, the term ``complete monogamy'' of the unified GlMEM was proposed. For a GlMEM $E^{(n)}$ that is coarsening monotonic (i.e., it satisfies Eq.~\eqref{coarsen} for any $\gamma\succ^a \gamma'$), it is said to be {{completely monogamous}} if for any
	$\rho\in\mathcal{S}^{A_1A_2\cdots A_n}$ that satisfies~\cite{Guo2020pra}
	\beax\label{cond2}
	E^{(k)}(\gamma)= E^{(l)}(\gamma')
	\eeax
	with $\gamma\succ^a \gamma'$ we have that
	\beax\label{cond2x}
	E^{(\ast)}({\gamma}) =0
	\eeax
	holds for all $\gamma\in \Xi(\gamma- \gamma')$, hereafter the superscript $(\ast)$ is associated with the partition $\gamma$, e.g., if $\gamma$ is an $m$-partite partition, then $(\ast)=(m)$. 
	For example, $E^{(3)}$ is completely monogamous if for any $\rho^{ABC}$ that admits $E^{(3)}(ABC)=E^{(2)}(AB)$ we get $E^{(2)}(AC)=E^{(2)}(BC)=0$. Let $E^{(n)}$ be a GlMEM that is tightly coarsening monotonic (i.e., it satisfies Eq.~\eqref{coarsen} for any $\gamma\succ^b \gamma'$). $E^{(n)}$ is defined to be
	{tightly complete monogamous} if for any $\rho\in\mathcal{S}^{A_1A_2\cdots A_n}$ that satisfies~\cite{Guo2020pra}
	\beax\label{cond3}
	E^{(k)}(\gamma)= E^{(l)}(\gamma')
	\eeax
	with $\gamma\succ^b \gamma'$ we have that
	\beax\label{cond3x}
	E^{(\ast)}({\gamma}) =0
	\eeax
	holds for all $\gamma\in \Xi(\gamma- \gamma')$. For instance, $E^{(3)}$ is tightly complete monogamous if for any $\rho^{ABC}$ that admits $E^{(3)}(ABC)=E^{(2)}(A|BC)$ we have $E^{(2)}(BC)=0$.


\section{The unified-$(q,s)$ entropy entanglement}


It was shown in Ref.~\cite{Hu2006jmp} that, $G_x^y(\rho)=\left( \tr\rho^x\right)^y $ is convex for $x\geq 1$, $xy\geq 1$ or $0<x<1$, $y<0$. It flows that
\bea\label{bipartite pure2}
E_{q,s}(|\psi\ra):=S_{q,s}(\rho^A),\quad q>1, \, qs\geq 1
\eea
for pure states, $\rho^A=\tr_B|\psi\ra\la\psi|$, and its convex-roof extension for mixed states,
is also a well-defined entanglement monotone since $S_{q,s}(\rho)$ is concave whenever $q>1$ and $qs\geq 1$.
Hereafter, we give only the measures for pure states, for the case of mixed states they are all defined by the convex-roof extension with no further statement. 
The reduced function $S_{q,s}(\rho)$ is strictly concave whenever $s\geq 1$ since
\beax 
\tr \left( \sum_ip_i\rho_i\right) ^q\leq\sum_ip_i\tr\rho_i^q 
\eeax
implies
\beax 
\left[ \tr\left( \sum_ip_i\rho_i\right) ^q\right]^{s}\leq\left( \sum_ip_i\tr\rho_i^q \right)^s\leq \sum_ip_i\left( \tr\rho_i^q \right)^s
\eeax
and $\tr\rho^q$ is strictly convex for $q>1$. The last inequality holds because $x^s$ is convex whenever $s\geq 1$. Moreover, we have the following lemma with the proof is given in Appendix~1.

\begin{lemma}\label{lemma1}
	For $q>1$,
	\begin{eqnarray}
		S_{q,\frac1q}(\rho^A)=\frac{q}{q-1}\left[1-(\tr\rho_A^q)^{1/q}\right] 
	\end{eqnarray}
	is strictly concave. 
\end{lemma}

By Lemma~\ref{lemma1}, $S_{q,s}$ is strictly concave whenever $qs\geq 1$. Thus,
according to Theorem 2 in Ref.~\cite{Vidal2000} and Theorem in Ref.~\cite{Guo2019pra}, together with the fact that $E_F$ is always convex, we can obtain the following result.

\begin{theorem}\label{thm:bipartite}
	$E_{q,s}$ is a continuous bipartite entanglement monotone and it is monogamous.
\end{theorem}

Note here that, $E_{r,t}$ is also monogamous since $g(\rho)=\tr(\rho^r)$ is also a strictly concave function. Going further, we compare $E_{q,s}$ with $E_{r,t}$. In general, for different parameters, they are incomparable. But we can obtain the the following relation (the proof is presented in Appendix~2).

\begin{pro}\label{pro1}
	For any bipartite entangled state $\rho$, 
	\bea \label{eq-pro1}
	E_{q,s}(\rho)<E_{q,1/s}(\rho)<E_{1/q,1/s}(\rho)
	\eea  
	whenever $q\geq s> 1$.
\end{pro}

We illustrate Eq.~\eqref{eq-pro1} with the following example for the two-qubit state.

\begin{example}
	The two-qubit Werner state is defined by
	\begin{eqnarray}
		\rho_{w}(p)=p|\Phi^+\rangle\langle\Phi^+|+\frac{1-p}{4}I_{4},
	\end{eqnarray}
	where $p\in[0,1]$, $|\Phi^+\rangle=\frac{1}{\sqrt2}(|00\ra+|11\ra)$.
\end{example}	

\begin{figure}[htp] 
	\centering              
	\includegraphics[width=0.4\textwidth]{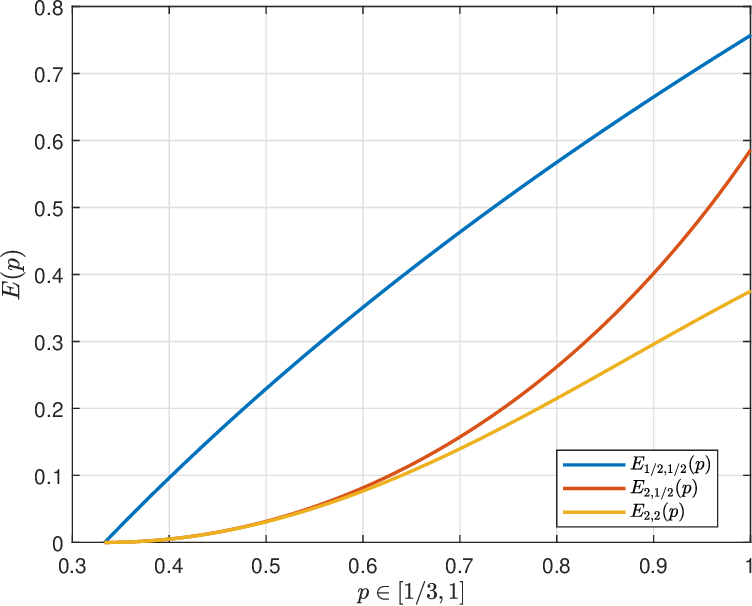} 
	\caption{The comparison of $E_{2,2}(\rho_{w})$, $E_{2,1/2}(\rho_{w})$ and $E_{1/2,1/2}(\rho_{w})$ for the Werner state $\rho_w(p)$ with $p\in[1/3,1]$}.   
	\label{fig1}	            
\end{figure}

Recalling that, for a bipartite pure two-qubit state $|\psi\rangle\in\mathcal{H}^{AB}$ with the Schmidt coefficients $\sqrt{\lambda},\sqrt{1-\lambda}$, the reduced function $h(\rho^A)$ can be denoted by $\check{h}(\lambda)$ for some $\check{h}$~\cite{Wootters1998prl}, i.e., $h(\rho^A)=\check{h}(\lambda)=E(|\psi\ra)$, where  $\check{h}(\lambda)=\check{h}(1-\lambda)$. 
According to the arguments in Ref.~\cite{Wootters1998prl}, for any convex-roof extended entanglement measure $E_F$ and any two-qubit mixed state $\rho$,
\begin{eqnarray}
	E_{F}(\rho)=\varepsilon\left[ C(\rho)\right] =\check{h}\left( \frac{1+\sqrt{1-C(\rho)^2}}{2}\right),
\end{eqnarray}
where $C$ is the concurrence which is defined by $C(|\psi\ra)=\sqrt{1-\tr\rho_A^2}$, $\varepsilon(C)$ is a convex function which is monotonically increasing and ranges from $0$ to $1$ whenever $C$ goes from $0$ to $1$.
By the definitions of $E_{q,s}$ and $E_{r,t}$, we can get
\bea
\check{h}_{q,s}(\lambda)&=&\frac{1}{(q-1)s}\left\lbrace 1-\left[\lambda^q+(1-\lambda)^q\right]^s\right\rbrace, \label{hqs}\\
\check{h}_{r,t}(\lambda)&=&\frac{1}{(1-r)t}\left\lbrace\left[\lambda^r+(1-\lambda)^r\right]^t-1\right\rbrace.\label{hrt}
\eea

Note that
\begin{widetext}
\bea
 E_{q,s}(\rho_w)&=&\varepsilon_{q,s}\left[ C(\rho_{w})\right]
 =\frac{1}{(q-1)s}\left\lbrace 1-\left[ \left( \frac{1+\sqrt{1-C(\rho_{w})^2}}{2}\right) ^q+\left( \frac{1-\sqrt{1-C(\rho_{w})^2}}{2}\right)^q\right] ^s\right\rbrace, 
\eea
\bea 
 E_{r,t}(\rho_w)&=&\varepsilon_{r,t}\left[ C(\rho_{w})\right]
 =\frac{1}{(1-r)t}\left\lbrace \left[ \left( \frac{1+\sqrt{1-C(\rho_{w})^2}}{2}\right) ^r+\left( \frac{1-\sqrt{1-C(\rho_{w})^2}}{2}\right)^r\right] ^t-1\right\rbrace,
\eea
\end{widetext}

where $\varepsilon_{q,s}\left[C(\rho_{w})\right]$ and $\varepsilon_{r,t}\left[C(\rho_{w})\right]$ are convex functions which are monotonically increasing  and ranges from $0$ to $1$ when $C(\rho_{w})$ goes from $0$ to $1$. The concurrence of such a Werner state has a very well-known exact analytical solution~\cite{Chen2006rmp}
\begin{eqnarray}
	C(\rho_{w})=\max\left\lbrace 0,{(3p-1)}/{2}\right\rbrace .
\end{eqnarray}
Let $q=s=2$. We have
\beax
E_{2,2}(\rho_{w})&=\begin{cases}
	\frac{(3p-1)^2}{8}-\frac{(3p-1)^4}{128},&~ p>1/3\\
	0,&~p\le1/3,
\end{cases}
\eeax 
\beax 
E_{1/2,1/2}(\rho_{w})&= \begin{cases}
	4\left[\left(\frac{3p+1}{2}\right)^{1/4}-1\right],&~ p>1/3\\
	0,&~p\le1/3,
\end{cases}
\eeax 
\beax 
E_{2,1/2}(\rho_{w})&=\begin{cases}
	2\left( 1-\sqrt{\frac{-9p^2+6p+7}{8}}\right),&~ p>1/3\\
	0,&~p\le1/3.
\end{cases}			
\eeax
The comparing of $E_{2,2}(\rho_{w})$, $E_{1/2,1/2}(\rho_{w})$, and $E_{2,1/2}(\rho_{w})$ is plotted in Fig.~\ref{fig1}. It is clear that $E_{2,2}(\rho_{w})<E_{2,1/2}(\rho_{w})<E_{1/2,1/2}(\rho_{w})$.

\begin{figure}[htp] 
\centering
	                       
\includegraphics[width=0.4\textwidth]{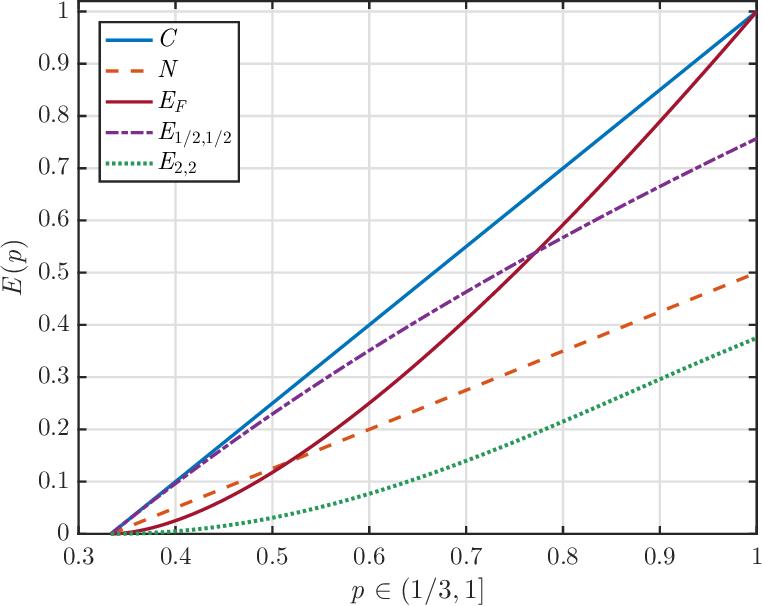} 
\caption{The comparison of $C$, $N$, $E_F$, $E_{2,2}$ and $E_{1/2,1/2}$ for the Werner state $\rho_w(p)$ with $p\in[1/3,1]$. } 
\label{fig2}  
\end{figure}

At last we compare the unified entropy entanglement with several well-known bipartite entanglement measures, such as concurrence $C$~\cite{Hill1997prl}, negativity $N$~\cite{Vidal2002pra}, and entanglement of formation $E_F$~\cite{Bennett1996pra2}. We recall that
\beax
C(|\psi\ra)=\sqrt{2(1-\tr\rho_A^2)},\\
E_f(|\psi\ra)=S(\rho_A),
\eeax
and $N$ is defined by 
\beax 
N(\rho)=\frac12(\|\rho^{T_A}\|_{\tr}-1)
\eeax
for both pure states and mixed states, where $\rho^{T_A}$ denotes the partial transpose of $\rho$ up to part $A$. We consider the two-qubit Werner state 
\bea 
\rho_w(p)=p|\Phi^+\rangle\langle\Phi^+|
+(1-p)I_4/4, \quad p\in[0,1],
\eea 
where
$|\Phi^+\rangle=(|00\rangle+|11\rangle)/\sqrt{2}$ is a
Bell state.
We take $c=\max\{0,(3p-1)/2\}$ and
$h_2(x)=-x\log_2 x-(1-x)\log_2(1-x)$ with $0\log_2 0=0$.
For the case of $q=s=2$ and $r=t=1/2$, the comparing of $C$, $N$, $E_F$, $E_{2,2}$ and $E_{1/2,1/2}$ for the Werner state $\rho_w(p)$ is plotted in Fig.~\ref{fig2}. It is clear that $C$ is always the largest or joint largest measure, while $E_{2,2}$ is always the smallest. Let $p_1$ and $p_2$
denote the two intersection points such that $E_F=N$ and
$E_F=E_{1/2,1/2}$, respectively. It is clear from the figure that
\beax
\begin{aligned}
	& C>E_{1/2,1/2}>N>E_F>E_{2,2} ~\textrm{whenever} ~\tfrac{1}{3}<p<p_1,\\
	& C>E_{1/2,1/2}>E_F>N>E_{2,2} ~\textrm{whenever} ~ p_1<p<p_2,\\
	& C>E_F>E_{1/2,1/2}>N>E_{2,2} ~\textrm{whenever} ~p_2<p<1.
\end{aligned}
\eeax
If $p=1$, i.e., for the Bell state $|\Phi^+\rangle$, we have $C=E_F>E_{1/2,1/2}>N>E_{2,2}$.

\if false 
\begin{table}[h]
	\caption{\label{tab:0} Comparison of concurrence $C$, negativity $N$, entanglement of formation $E_f$, $E_{q,s}$, and $E_{r,t}$ for $\rho_w(p)$ and $|\Phi^+\rangle$. }	
	\begin{indented}
		\item[]		
		\begin{tabular}{@{}lll}
			\br
			E &  $\rho_w(p)$  & $|\Phi^+\rangle$ \\
			\mr
			$C$ & $c$ & 1  \\
			$N$ & $c/2$ & 1/2  \\			
			$E_F$ & $h_2(\frac{1+\sqrt{1-c^2}}{2})$ & 1  \\			
			$E_{q,s}$ & $\check{h}_{q,s}(\frac{1+\sqrt{1-c^2}}{2})$ & $\dfrac{1-2^{(1-q)s}}{(q-1)s}$ \\			
			$E_{r,t}$ & $\check{h}_{r,t}(\frac{1+\sqrt{1-c^2}}{2})$ & $\dfrac{2^{(1-r)t}-1}{(1-r)t}$ \\								
			\br
		\end{tabular}
	\end{indented}
\end{table}
\fi


\section{Global MEM induced by the unified entropy}


A natural candidate for the global multipartite entanglement measure (GlMEM) is the one defined by the sum of the reduced functions on all the single subsystems~\cite{Guo2020pra,Guo2024rip}, i.e.,
\bea\label{sum1}
E^{(n)}(|\psi\ra):=\frac{1}{2}\sum_ih(\rho^{A_i}),
\eea
where $\rho^{X}=\tr_{\overline{X}}|\psi\ra\la\psi|$.
For any system that is split into two parts $XY$, $\overline{X}$ denotes the complementary part of $X$, i.e., $\overline{X}=Y$. For example, in the system $ABCD$, $\overline{A}=BCD$, $\overline{AB}=CD$, etc. For the mixed states, it is defined as in Eq.~\eqref{convex-roof}.

For any pure state $|\psi\rangle\in\mH^{A_1A_2\cdots A_n}$, we define
\bea\label{multipartite pure}
E_{q,s}^{(n)}(|\psi\rangle):=\frac{1}{2(q-1)s}\left[n-\sum_{j=1}^{n}\left( \tr\rho_{A_j}^q\right) ^s\right],
\eea
where $\rho^{X}=\tr_{\overline{\textsc{X}}}|\psi\rangle\la\psi|$. By definition, $E_{q,s}^{(n)}$ is a unified GlMEM by Theorem 1 in Ref.~\cite{Guo2024rip}. In addition, according to Theorem 1 in Ref.~\cite{Guo2024rip}, $E_{q,s}^{(n)}$ is completely monogamous, and if (i) the reduced function 
\bea \label{h_qs}
h_{q,s}(\rho)=\frac{1}{(q-1)s}\left[  1-(\tr\rho^q)^s\right]
\eea is subadditive [$h$ is called subadditive if $h(\rho^{AB})\leq h(\rho^A)+h(\rho^B)$ holds for any $\rho^{AB}$], it is a complete GlMEM, and if (ii) $h_{q,s}(\rho)$ is subadditive with 
\be\label{hierachy-h}
h_{q,s}(\rho^{AB})=h_{q,s}(\rho^A)+h_{q,s}(\rho^B)\Rightarrow\rho^{AB}~\text{is separable}, 
\ee
it is tightly complete monogamous. As we will show below (the detailed proof is given in Appendix~3), $h_{q,s}(\rho)$ is subadditive and admits the condition in Eq.~\eqref{hierachy-h}.

\begin{lemma}\label{lemma2}
	For any $\rho_{AB}\in\mS^{AB}$, $q>1$ and $qs\geqslant 1$, 
	\begin{eqnarray}\label{lemma tr}
		1+\left(\tr\rho_{AB}^{q}\right)^{s}\ge\left(\tr\rho_A^{q}\right)^{s}+\left(\tr\rho_B^{q}\right)^{s}, 
	\end{eqnarray}
	and the equality holds if and only if $\rho^{A}$ or $\rho^{B}$ is a pure state.
\end{lemma}

By Lemma~\ref{lemma2} and Theorem 1 in~\cite{Guo2024rip}, we have the following conclusion.

\begin{theorem}\label{th:tripartite2}
	$E_{q,s}^{(n)}$ is a complete GlMEMo and it is not only completely monogamous but also tightly completely monogamous.
\end{theorem}

Here ``Mo'' in ``*Mo'' refers to it is a ``* monotone''.
Similarly, we can define
\bea\label{multipartite pure'}
E_{r,t}^{(n)}(|\psi\rangle):=\frac{1}{2(1-r)t}\left[\sum_{j=1}^{n}\left( \tr\rho_{A_j}^r\right) ^t-n\right].
\eea
By Theorem 1 in~\cite{Guo2024rip}, it is a unified GlMEM and it is completely monogamous. However its reduced function $h_{r,t}(\rho)=\frac{1}{(r-1)t}\left[1-(\tr\rho^r)^t\right]$ is not subadditive, i.e.,
\begin{eqnarray}\label{counter-ex1}
	1+\left(\tr\rho_{AB}^r\right)^{t}>\left(\tr\rho_A^{r}\right)^{t}+\left(\tr\rho_B^{r}\right)^{t}
\end{eqnarray}
for some $\rho^{AB}\in\mS^{AB}$. So $E_{r,t}^{(n)}$ is not a complete MEM, and it is not tightly complete monogamous. For example, we take $\rho^{AB}=\rho^{A}\otimes\rho^{B}$, where the eigenvalues of $\rho^{A}$ coincide with that of $\rho^B$ and they are not pure states. It turns out that
$\tr\rho_{AB}^{r}=\tr(\rho_{A}\otimes\rho_{B})^{r}=\tr\left(  \rho_{A}^r\ot\rho_{A}^r\right) 
=\left(\tr\rho_{A}^r\right)^2$. Let $x=\left(  \tr\rho_A^{r}\right) ^{t}$, then $x>1$ and $1+x^2>2x$, that is, Eq.~\eqref{counter-ex1} holds for such a state. We compare $E_{q,s}^{(n)}$ and $E_{r,t}^{(n)}$ in Table~\ref{tab:1} for more clarity.

\begin{table}[htp]
	\caption{\label{tab:1} Comparison of  $E_{q,s}^{(n)}$ and $E_{r,t}^{(n)}$. CM is the abbreviation for ``completely monogamous'', TCM is the abbreviation for ``tightly completely monogamous''.}	
	\begin{ruledtabular}	
		\begin{tabular}{@{}llllll}
			$E^{(n)}$     &  Convex &  Unified & Complete & CM & TCM \\
			\midrule
			$E_{q,s}^{(n)}$ & $\checkmark$ & $\checkmark$ & $\checkmark$ & $\checkmark$ & $\checkmark$ \\
			$E_{r,t}^{(n)}$  & $\checkmark$ & $\checkmark$ & $\times$ & $\checkmark$ & $\times$\\ 										
		\end{tabular}		
\end{ruledtabular}
\end{table}

According to the proof of the Proposition~\ref{pro1}, we can obtain the following proposition.

\begin{pro}
	For any $n$-partite entangled state $\rho$, 
	\bea\label{eq-pro2} 
	E_{q,s}^{(n)}(\rho)<E_{q,1/s}^{(n)}(\rho)<E_{1/q,1/s}^{(n)}(\rho)
	\eea  
	whenever $q\geq s> 1$.
\end{pro}

We illustrate Eq.~\eqref{eq-pro2} with the following three-qubit states. For the three-qubit GHZ state $|\mathrm{GHZ}_3\rangle=\frac12(|000\rangle+|111\rangle)$, it is easy to see that
\beax
E_{q,s}^{(3)}\big(|\mathrm{GHZ}_3\rangle\big)&=&\frac{3\big[1-2^{s(1-q)}\big]}{2(q-1)s},\\
E_{q,1/s}^{(3)}\big(|\mathrm{GHZ}_3\rangle\big)&=&\frac{3s\big[1-2^{\frac{(1-q)}{s}}\big]}{2(q-1)},\\ E_{1/q,1/s}^{(3)}\big(|\mathrm{GHZ}_3\rangle\big)&=&\frac{3sq\left( 1-2^{\frac{q-1}{qs}}\right) }{2(q-1)},
\eeax
and thus
\beax
E_{q,s}^{(3)}\big(|\mathrm{GHZ}_3\rangle\big)<E_{q,1/s}^{(3)}\big(|\mathrm{GHZ}_3\rangle\big)<E_{1/q,1/s}^{(3)}\big(|\mathrm{GHZ}_3\rangle\big).
\eeax
For the three-qubit W state $|\mathrm{W}\rangle=\frac13(|001\rangle+|010\rangle+|100\rangle)$, we have
\beax
E_{q,s}^{(3)}\big(|\mathrm{W}\rangle\big)&=&\frac{3\left\lbrace 1-[(\frac{2}{3})^q+(\frac{1}{3})^q]^s\right\rbrace }{2(q-1)s},\\
E_{q,1/s}^{(3)}\big(|\mathrm{W}\rangle\big) &=&\frac{3s\left\lbrace 1-[(\frac{2}{3})^q+(\frac{1}{3})^q]^{1/s}\right\rbrace }{2(q-1)},\\
E_{1/q,1/s}^{(3)}\big(|\mathrm{W}\rangle\big)& =&\frac{3sq\left\lbrace 1-[(\frac{2}{3})^{1/q}+(\frac{1}{3})^{1/q}]^{1/s}\right\rbrace }{2(q-1)}.
\eeax
We thus get
\beax
E_{q,s}^{(3)}\big(|\mathrm{W}\rangle\big)<E_{q,1/s}^{(3)}\big(|\mathrm{W}\rangle\big)<E_{1/q,1/s}^{(3)}\big(|\mathrm{W}\rangle\big).
\eeax

\begin{table}[htp]
	\caption{\label{tab:2} Comparison of different tripartite entanglement measures for GHZ state and W state.}	
\begin{ruledtabular}	
		\begin{tabular}{@{}lll}
			$E^{(3)}$ & $|\mathrm{GHZ}\rangle$ & $|\mathrm{W}\rangle$ \\
			\midrule
			$C^{(3)}$ & $\sqrt{3/2}$ & $2/\sqrt{3}$ \\			
			$\tau^{(3)}$ & $3/2$ & $4/3$ \\			
			$N^{(3)}$ & 3 & $2\sqrt{2}$  \\			
			$E_{q,s}^{(3)}$ & $\frac{3\big[1-2^{s(1-q)}\big]}{2(q-1)s}$ &   $\frac{3\left\lbrace 1-[(\frac{2}{3})^q+(\frac{1}{3})^q]^s\right\rbrace }{2(q-1)s}$ \\			
			$E_{r,t}^{(3)}$  & $\frac{3\big[2^{t(1-r)}-1\big]}{2(1-r)t}$ & $\frac{3\left\lbrace [(\frac{2}{3})^r+(\frac{1}{3})^r]^t-1\right\rbrace }{2(1-r)t}$  \\	
			$E_{2,2}^{(3)}$ & $9/16$ &   $14/27$ \\	
			$E_{2,1}^{(3)}$  & $3/4$ & $2/3$  \\	
			$E_{3,1}^{(3)}$  & $9/16$ & $1/2$  \\	
			$E_{1/2,1/2}^{(3)}$ & $6(\sqrt[4]{2}-1)$ &  $6 \sqrt{\frac{\sqrt{2}+1}{\sqrt{3}}-1} $ \\	
			$E_{1/4,1/2}^{(3)}$ & $4(\sqrt[8]{8}-1)$ &  $4 \sqrt{\frac{\sqrt[4]{2}+1}{\sqrt[4]{3}}-1} $ \\						
		\end{tabular}		
\end{ruledtabular}
\end{table} 

We now compare the tripartite concurrence $C^{(3)}$~\cite{Guo2020pra}, tripartite negativity $N^{(3)}$~\cite{Guo2020pra}, $E_{r,t}^{(3)}$, and $E_{r,t}^{(3)}$ for 
$|\mathrm{GHZ}\rangle=(|000\rangle+|111\rangle)/\sqrt{2}$
and
$|\mathrm{W}\rangle=(|001\rangle+|010\rangle+|100\rangle)/\sqrt{3}$.
The tripartite negativity is defined as~\cite{Guo2020pra}
$N^{(3)}(\rho)
=\sum_{j=A,B,C}\|\rho^{T_j}\|_{\tr}-3$,
where $T_j$ denotes partial transposition with respect to subsystem $j$.


As listed in Table~\ref{tab:2}, all these measures assign a larger value to the GHZ state than that of the W state and  $E^{(3)}_{2,2}<E^{(3)}_{2,1}<E^{(3)}_{1/2,1/2}<C^{(3)}<\tau^{(3)}<N^{(3)}$ for both states.
On the other hand, the ordering of different measures depend on the state: $E^{(3)}_{1/4,1/2}<C^{(3)}$ for the GHZ state, whereas $E^{(3)}_{1/4,1/2}>C^{(3)}$ for the W state.
Moreover, $E^{(3)}_{2,2}=E^{(3)}_{3,1}$ for the GHZ state, while $E^{(3)}_{2,2}>E^{(3)}_{3,1}$ for the W state.

	
	\section{GMEMs induced by the unified entropy}
	

	Recall that, a multipartite state $|\psi\ra$ is genuinely entangled if it is entangled under any bipartition, and a mixed state is genuinely entangled if it cannot be represented as a convex combination of biseparable pure states. Here, $|\psi\ra$ is called biseparable if it is separable under some bipartition. An MEM $E_g^{(n)}: \mS^{A_1A_2\cdots A_n}\to\bb{R}^{+}$ is defined to be a genuine entanglement measure (GEM) if (i) $E_g^{(n)}(\rho)=0$ for any biseparable  $\rho$ and 
	$E_g^{(n)}$ is convex. 
	For any reduced function $h$, 
	\bea\label{gmin2}
	E_{g}^{(n)}(|\psi\ra)
	:=\min_{\gamma_i\in\Gamma_2}h(\rho^{{X_{1(i)}}})
	\eea
	is a GEMo~\cite{Guo2024rip}. For example, GMC, denoted by $C_{gme}$~\cite{Ma2011pra}, is defined as in Eq.~\eqref{gmin2}. Recall that,
	\bea\label{Cgme}
	C_{gme}(|\psi\ra):=\min\limits_{\gamma_i \in \Gamma_2} \sqrt{2\left( 1-\tr\rho_{X_{1(i)}}^{2}\right) }
	\eea
	for pure state $|\psi\ra\in\mH^{A_1A_2\cdots A_n}$. 
	
	For an $n$-partite pure state $|\psi\rangle\in\mH^{A_1A_2\cdots A_n}$, we define GEMos induced by the unified entropy, i.e.,
	\begin{eqnarray}
		E_{g(q,s)}^{(n)}(|\psi\ra)&:=&\min\limits_{\gamma_i \in \Gamma_2}\left\lbrace \frac{1}{(q-1)s}\left[ 1-\left( \tr\rho_{{X_{1(i)}}}^{q}\right) ^{s}\right] \right\rbrace ,~~~~\label{GME definition}\\
		E_{g(r,t)}^{(n)}(|\psi\ra)&:=&\min\limits_{\gamma_i\in\Gamma_2}\left\lbrace \frac{1}{(1-r)t}\left[\left( \tr\rho_{{X_{1(i)}}}^{r}\right)^{t}-1\right]\right\rbrace.~~~~\label{GME definition2}
	\end{eqnarray}

	We now compare $E_{g(q,s)}^{(3)}$, $E_{g(r,t)}^{(3)}$, and $C_{gme}$ for GHZ and W states in Table~\ref{tab:3}. It is clear that each measure assigns a larger value to $|\mathrm{GHZ}\rangle$ than to $|\mathrm{W}\rangle$ as desired. For the special choices of parameters choices, 
	\begin{widetext}
	\bea
	\begin{aligned}
		|\mathrm{GHZ}\ra:\quad
		&E_{g(2,1)}^{(3)}
		=E_{g(3,1/2)}^{(3)}
		<E_{g(1/2,1/2)}^{(3)}
		<E_{g(1/2,2/3)}^{(3)}
		<C_{\mathrm{gme}},\\
		|\mathrm{W}\ra:\quad
		&E_{g(3,1/2)}^{(3)}
		<E_{g(2,1)}^{(3)}
		<E_{g(1/2,1/2)}^{(3)}
		<E_{g(1/2,2/3)}^{(3)}
		<C_{\mathrm{gme}}.
	\end{aligned}
	\eea
	\end{widetext}

	\begin{table}[htp]
		\caption{\label{tab:3} Comparison of $E_{g(q,s)}^{(3)}$, $E_{g(r,t)}^{(3)}$, and $C_{gme}$ for GHZ state and W state.}	
	\begin{ruledtabular}
			\begin{tabular}{@{}lll}
				$E_g^{(3)}$ & $|\mathrm{GHZ}\rangle$ & $|\mathrm{W}\rangle$ \\
				\midrule
				$C_{gme}$ & 1 & $\frac{2\sqrt{2}}{3}$ \\			
				$E_{g(q,s)}^{(3)}$ & $\frac{\big[1-2^{s(1-q)}\big]}{(q-1)s}$ &   $\frac{ 1-[(\frac{2}{3})^q+(\frac{1}{3})^q]^s }{(q-1)s}$ \\			
				$E_{g(r,t)}^{(3)}$  & $\frac{\big[2^{t(1-r)}-1\big]}{(1-r)t}$ & $\frac{ [(\frac{2}{3})^r+(\frac{1}{3})^r]^t-1 }{(1-r)t}$  \\			
				$E_{g(2,1)}^{(3)}$ & $\frac{1}{2}$ & $\frac{4}{9}$ \\			
				$E_{g(3,1/2)}^{(3)}$ & $\frac{1}{2}$ & $1-\frac{1}{\sqrt{3}}$ \\			
				$E_{g(1/2,1/2)}^{(3)}$ & $4(\sqrt[4]{2}-1)$ & $4(\sqrt[4]{\frac{3+2\sqrt{2}}{3}}-1)$ \\			
				$E_{g(1/2,2/3)}^{(3)}$ & $3(\sqrt[3]{2}-1)$ & $3(\sqrt[3]{\frac{3+2\sqrt{2}}{3}}-1)$ \\					
			\end{tabular}		
	\end{ruledtabular}
	\end{table}

	The calculation of $E_{g}^{(n)}$ is difficult for mixed states since the optimization procedure is always required. But the relation between $E_{g}^{(n)}$ and the bipartite entanglement or the global entanglement in the state can be established~\cite{Li2024qip}. For example, the relationship between $C_{gme}$ and the bipartite concurrences, and the relation between $C_{gme}$ and the global concurrence were obtained~\cite{Li2024qip}. In this section, we explore such an issue for these measures induced by the unified entropy we defined above.

	For simplicity, we fix some notations which are necessary in the flowing discussion.
	For a given $d$-dimensional quantum system,
	we let $c_{d}:=\min\limits_{\rho}(\tr\rho^{q})^{s}$, 
	where the minimum is taken over all qudit states, then $c_{d}=1/d^{(q-1)s}$. In order to see this, we
	let $\lambda_k^{2}$ be the eigenvalues of $\rho$, $s>0$.
	From the power mean inequality, when $q>1$, 
	$\dfrac{(\lambda_1^2)^q+(\lambda_2^2)^q+\cdots+(\lambda_d^2)^q}{d}\geq\left( \frac{\lambda_1^2+\lambda_2^2+\cdots+\lambda_d^2}{d}\right)^{q}
	=\dfrac{1}{d^{q}}$.
	Thus,
	$(\tr\rho^q)^s\geq1/{d^{(q-1)s}}$, and the equality holds if and only if $\rho ={I_{d}}/{d}$. Similarly, we let $l_{d}:=\max\limits_{\rho}(\tr\rho^{r})^{t}=d^{(1-r)t}$.
	Hereafter, for a given entanglement measure $E$, $E^{\gamma}$ denotes the entanglement is quantified under the partition $\gamma$ of the given system. We now give the following propositions with the proofs are presented in Appendix~4 and~Appendix 5, respectively.


	\begin{pro}\label{Lower bipartiten}
		Let $\rho\in\mS^{A_1A_2\cdots A_n}$ be an $n$-qudit state. Then
		\bea
		E_{g(q,s)}^{(n)}(\rho)&\ge&\sum\limits_{\gamma_i \in \Gamma_2}E_{q,s}^{\gamma_i}(\rho)+\dfrac{(2^{n-1}-2)(c_{d}-1)}{(q-1)s},\label{lower1-1}\\
		E_{g(r,t)}^{(n)}(\rho)&\ge&\sum\limits_{\gamma_i \in \Gamma_2}E_{r,t}^{\gamma_i}(\rho)+\dfrac{(2^{n-1}-2)(l_{d}-1)}{(r-1)t}.\label{lower1-2}
		\eea
	\end{pro}
	
	\begin{pro}\label{Lower multipartiten}
		Let $\rho\in\mS^{A_1A_2\cdots A_n}$ be an $n$-qudit state. Then
		\bea
		E_{g(q,s)}^{(n)}(\rho)&\ge&2E_{q,s}^{(n)}(\rho)+\dfrac{(n-1)(c_{d}-1)}{(q-1)s},\label{lower2-1}\\
		E_{g(r,t)}^{(n)}(\rho)&\ge&2E_{r,t}^{(n)}(\rho)+\dfrac{(n-1)(l_{d}-1)}{(r-1)t}.\label{lower2-2}
		\eea
	\end{pro}
	
	The lower bounds in Propositions~\ref{Lower bipartiten} and \ref{Lower multipartiten} are not tight in general.

	
	\section{Conclusion}
	

	Here we have presented entanglement monotones via the unified entropy. The multipartite system was also discussed. We found that, under the theory of the complete MEM, the two kinds of unified entropy are far different from each other when they are used for quantifying global multipartite entanglement since the reduced function $h_{q,s}$ is subadditive but $h_{r,t}$ is not. Moreover, we discussed the monogamy and the complete monogamy of these measures. We proved that all the unified entropy induced bipartite measures are monogamous since the reduced functions are strictly concave and that all the associated GlMEMs are completely monogamous. In particular, the  $(q,s)$-type GlMEM via the summation of the reduced functions is a complete GlMEMo and it is also tightly complete monogamous but the  $(r,t)$-type one is not. This suggests that the unified-$(q,s)$ entropy is better than the unified-$(r,t)$ entropy in such a sense. Together with the related results in literature, we now obtained a whole theory of the unified entropy based entanglement measure.


	\begin{acknowledgements}
		Y.G is supported by the National Natural Science Foundation of China under Grant Nos.~12471434 and 11971277, the Program for Young Talents of Science and Technology in Universities of Inner Mongolia Autonomous Region under Grant No. NJYT25010, the High-Level Talent Research Start-up Fund of Inner Mongolia University under Grant No. 10000-A260015/501, and the Inner Mongolia Autonomous Region Science and Technology Plan Projects under Grant No. 2025KYPT0098. S. D is supported by the National Natural Science Foundation of China under Grant No.~12271452.
	\end{acknowledgements}

	
	\appendix
	
	
	\section*{Appendix 1: The proof of Lemma~\ref{lemma1}}\label{appendix-A}

	For $q>1$, recall that the Schatten $q$-norm of an operator $X$ is defined by
	\begin{eqnarray}\notag
		\|X\|_q=\left(\operatorname{tr}|X|^q\right)^{1/q},
		\quad |X|=\sqrt{X^\dagger X}.
	\end{eqnarray}
	Since any state $\rho$ is positive,
	\begin{eqnarray}\notag
		\|\rho\|_q=\left[\operatorname{tr}(\rho)^q\right]^{1/q},
	\end{eqnarray}
	and consequently
	\begin{eqnarray}\notag
		S_{q,\frac{1}{q}}(\rho)=\frac{q}{q-1}\left(1-\|\rho\|_q\right).
	\end{eqnarray}
	
	For $q>1$, the Schatten $q$-norm is strictly convex. Thus, for any two
	distinct density operators $\rho$, $\sigma$, and any
	$\lambda\in(0,1)$,
	\begin{eqnarray}\notag
		\left\|\lambda\rho+(1-\lambda)\sigma\right\|_q<\lambda\|\rho\|_q+(1-\lambda)\|\sigma\|_q.	
	\end{eqnarray}
	It follows that
	\begin{eqnarray}\notag
		\begin{aligned}
			&S_{q,\frac{1}{q}}\bigl(\lambda\rho+(1-\lambda)\sigma\bigr)\\
			=&\frac{q}{q-1}\left[1-\left\|\lambda\rho+(1-\lambda)\sigma\right\|_q\right] \\
			>&\lambda S_{q,\frac{1}{q}}(\rho)+(1-\lambda)S_{q,\frac{1}{q}}(\sigma).
		\end{aligned}	
	\end{eqnarray}
	Hence, $S_{q,\frac{1}{q}}(\rho)$ is strictly concave for $q>1$.

	\section*{Appendix 2: The proof of Proposition~\ref{pro1}}\label{appendix-B}

	First, we consider the case of entangled pure state $|\psi\rangle\in\mH^{AB}$.
	
	(i) We first prove $E_{q,s}(|\psi\rangle)<E_{q,1/s}(|\psi\rangle)$ whenever $q\geq s> 1$.
	Let $\tr\rho^q=t$, where $\rho=\tr_B|\psi\ra\la\psi|$ and $f(t,s)=\frac{1}{s}(1-t^s)$. Then
	\begin{align*}
		E_{q,s}(|\psi\rangle)=\frac{1}{q-1}f(t,s), E_{q,1/s}(|\psi\rangle)=\frac{1}{q-1}f(t,1/s).
	\end{align*}
	So we only need to show $f(t,s)$ is strictly decreasing for any fixed $0<t<1$. Note that
	\beax
	\frac{\partial f(t,s)}{\partial s}=-\frac{1}{s^2}(1-t^s)-\frac{t^s\ln t}{s}.
	\eeax
	We let $y(s)=-st^s\ln t+t^s-1$. Then
	\begin{align*}
		y'(s)=-t^s\ln t-st^s(\ln t)^2+t^s\ln t=-st^s(\ln t)^2.
	\end{align*}
	It is clear that $y'(s)<0$ since $s>0$ and $t>0$.
	Since $y(0)=0$ and $y(s)$ is strictly decreasing, for any $s>0$, it always holds that $y(s)<0$ and thus $\frac{\partial f(t,s)}{\partial s}<0$. This implies that, for any fixed $t$, $f(t,s)$ is strictly decreasing with respect to $s$ when $s>0$.

	(ii) We now prove $E_{q,1/s}(|\psi\rangle)<E_{1/q,1/s}(|\psi\rangle)$ whenever $q\geq s> 1$. Since 
	\beax
	E_{q,1/s}(|\psi\rangle)=\dfrac{(1-[\tr(\rho^q)]^{1/s})s}{q-1}
	\eeax 
	and 
	\beax 
	E_{1/q,1/s}(|\psi\rangle)=\dfrac{([\tr(\rho^{1/q})]^{1/s}-1)s}{1-1/q},
	\eeax
	where $\rho=\tr_B|\psi\rangle\langle\psi|$, we only need to prove
	\begin{eqnarray}\label{eq-c}
		q([\tr(\rho^{1/q})]^{1/s}-1)>1-[\tr(\rho^q)]^{1/s}.
	\end{eqnarray}
	Since $q>1$, we let $x_{1/q}=\tr(\rho^{1/q})$ and $x_q=\tr(\rho^q)$. Then $x_{1/q}>1$ and $0<x_q<1$. We define 
	\begin{eqnarray}
		g(r)=q(x_{1/q}^r-1)+x_q^r-1, 
	\end{eqnarray}
	and $g(0)=0$. We get
	\begin{eqnarray*}
		g'(r)=qx_{1/q}^r\ln(x_{1/q})+x_q^r\ln(x_q)
	\end{eqnarray*}
	and
	\begin{eqnarray*}
		g'(0)=q\ln(x_{1/q})+\ln(x_q).
	\end{eqnarray*}
	
	Note that the quantum R\'{e}nyi entropy $S_{\alpha}(\rho)=\frac{1}{1-\alpha}\ln\tr(\rho^{\alpha})$, so $g'(0)$ can be expressed as
	\bea
	g'(0)=(q-1)[S_{1/q}(\rho)-S_{q}(\rho)].
	\eea
	Since $S_{\alpha}(\rho)$ strictly decreases as the parameter $\alpha$ increases (for mixed states) and $1/q<q$, we have $S_{1/q}(\rho)>S_{q}(\rho)$ and $g'(0)>0$.
	
	For any $r>0$, $x_{1/q}^r>1$ and $x_q^r<1$. In addition, $q\ln(x_{1/q})>0$ and $(-\ln(x_q))>0$, therefore
	\begin{eqnarray*}
		g'(r)&=&q\ln(x_{1/q})\cdot x_{1/q}^r-[-\ln(x_q)]\cdot x_q^r\\
		&>&q\ln(x_{1/q})\cdot1-[-\ln(x_q)]\cdot1\\
		&=&q\ln(x_{1/q})-[-\ln(x_q)]\\
		&=&g'(0)>0.
	\end{eqnarray*}
	This implies that $g(r)$ is a strictly increasing function when $r>0$, so it is certain that $g(r)>0$ for any $r>0$. When $r=1/s>0$, $g(1/s)>0$, i.e., Eq.~\eqref{eq-c} holds.

	We next consider the case of mixed states. 
	Let $\rho$ be an entangled mixed state in $\mS^{AB}$.
	We first prove that $E_{q,s}(\rho)<E_{q,1/s}(\rho)$ for $q\geq s>1$. 
	Let
	$\rho=\sum_i p_i|\psi_i\rangle\langle\psi_i|$ be an optimal pure-state decomposition so that
	\beax
	E_{q,1/s}(\rho)
	=\sum_i p_i E_{q,1/s}(|\psi_i\rangle).
	\eeax
	On the other hand,
	\beax
	E_{q,s}(\rho)
	\leq
	\sum_i p_i E_{q,s}(|\psi_i\rangle)
	\eeax
	since $E_{q,s}$ is convex. 
	By the result proved above for pure states, $E_{q,s}(|\psi_i\rangle)
	\leq E_{q,1/s}(|\psi_i\rangle)$ for every $i$, and the inequality is strict whenever $|\psi_i\rangle$ is entangled. Moreover, since $\rho$ is entangled, there exists at least some $i_0$ such that $|\psi_{i_0}\rangle$ is entangled. Therefore,
	\beax
	\sum_i p_i E_{q,s}(|\psi_i\rangle)<\sum_i p_i E_{q,1/s}(|\psi_i\rangle)=E_{q,1/s}(\rho),
	\eeax
	and thus
	\beax
	E_{q,s}(\rho)<E_{q,1/s}(\rho).
	\eeax
	
	To prove the second inequality, we let $\rho=\sum_j r_j|\phi_j\rangle\langle\phi_j|$ be an optimal pure-state decomposition such that $E_{1/q,1/s}(\rho)=\sum_j r_j E_{q,1/s}(|\phi_j\rangle)$. Then
	\begin{eqnarray}
		E_{q,1/s}(\rho)
		\leq
		\sum_j r_j E_{q,1/s}(|\phi_j\rangle).
	\end{eqnarray}
	Again, since $\rho$ is entangled, we obtain
	\begin{eqnarray}
		E_{q,1/s}(\rho)<\sum_j r_j E_{1/q,1/s}(|\phi_j\rangle)=E_{1/q,1/s}(\rho).
	\end{eqnarray}
	This completes the proof.

	\section*{Appendix 3: The proof of Lemma~\ref{lemma2}}\label{appendix-C}

	There are two cases: (i) $qs=1$ and (ii) $qs>1$.

	(i) If $qs=1$,  Eq.\eqref{lemma tr} reduces to Audenaert?s Schatten $q$-norm inequality
	\begin{eqnarray}\label{lemma tr 1}
		1+\left(\tr\rho_{AB}^{q}\right)^{1/q}\ge\left(\tr\rho_A^{q}\right)^{1/q}+\left(\tr\rho_B^{q}\right)^{1/q},
	\end{eqnarray}
	and equality is attained whenever either $\rho_A$ or $\rho_B$ is pure~\cite{Audenaert2007jmp}. We now determine the necessary condition for the equation to hold.
	
	Let
	\beax
	r=\frac{q}{q-1},
	\eeax
	then $\frac1q+\frac1r=1$ and $r>1$. We define
	\beax
	X= \frac{\rho_A^{\,q-1}} {\|\rho_A\|_q^{\,q-1}}, \quad Y= \frac{\rho_B^{\,q-1}} {\|\rho_B\|_q^{\,q-1}}.
	\eeax
	Then
	\beax
	\|X\|_r^r = \frac{\tr\rho_A^{(q-1)r}} {\|\rho_A\|_q^{(q-1)r}} = \frac{\tr\rho_A^q} {\|\rho_A\|_q^q} =1,
	\eeax
	and similarly $\|Y\|_r^{r}=1$. Moreover,
	\begin{equation}\label{lemma tr 3}
	\begin{split}
		\tr(X\rho_A)&=\frac{\tr\rho_A^{q}} {\|\rho_A\|_q^{q-1}}=\|\rho_A\|_q, \\ \tr(Y\rho_B)&=\frac{\tr\rho_B^{q}} {\|\rho_B\|_q^{q-1}}=\|\rho_B\|_q. 
	\end{split}
	\end{equation}
	
	
	We first show that if both $\rho_A$ and $\rho_B$ are mixed, equality in Eq.\eqref{lemma tr 1} can not occur. 
	We suppose therefore that $\rank\rho_A\ge2$ and $\rank\rho_B\ge2$. Let $\{x_i\}$ and $\{y_j\}$ denote the positive eigenvalues of $X$ and $Y$, respectively. Then $\sum_i x_i^r=\sum_j y_j^r=1$ since $\|X\|_r=\|Y\|_r=1$. On the other hand, we have
	\beax
	(x_i+y_j-1)_+<x_i y_j,
	\eeax
	where $(t)_+=\max\{t,0\}$. This reveals $(x_i+y_j-1)_+^r<x_i^r y_j^r$.
	Therefore,
	\beax
	\sum_{i,j}(x_i+y_j-1)_+^r<\sum_{i,j}x_i^r y_j^r=\left(\sum_i x_i^r\right) \left(\sum_jy_j^r\right)=1.
	\eeax
	We let $H=X\otimes I_B+I_A\otimes Y-I_{AB}$ and $\{|a_i\rangle\}$ be the orthonormal basis induced by the eigenvectors of $X$ and $\{|b_j\rangle\}$ be the orthonormal basis induced by eigenvectors of $Y$ with $Y|b_j\rangle = y_j|b_j\rangle$. 
	In the basis $\{|a_i\rangle|b_j\rangle\}$, we have
	\beax
	H|a_i\ra| b_j\rangle&=&(X\otimes I_B + I_A\otimes Y - I_{AB})|a_i b_j\rangle\\
	&=& (x_i+y_j-1)|a_i\ra| b_j\rangle.
	\eeax
	Therefore,  the eigenvalues of $H$ are precisely $x_i+y_j-1$.
	
	For the Jordan decomposition of $H$, $H = H_+ -H_-$, we have
	$\|H_+\|_r^r = \sum_{i,j}(x_i+y_j-1)_+^r<1$,
	and hence 
	\begin{eqnarray}\label{lemma tr 4}
		\|H_+\|_r<1.
	\end{eqnarray}
	By Eq.\eqref{lemma tr 3}, we can obtain
	\bea\label{lemma tr 5}
	\|\rho_A\|_q+\|\rho_B\|_q &=& \tr(X\rho_A) + \tr(Y\rho_B)\\\notag 
	&=&\tr \left[ (X\otimes I_B+I_A\otimes Y)\rho_{AB} \right]\\\notag  
	&=& \tr[(I_{AB}+H)\rho_{AB}]\\\notag 
	&=& 1+\tr(H\rho_{AB}).
	\eea
	Since $H_+-H=H_{-}\ge0$, we have $H\le H_+$, and thus $\tr\big[(H_+-H)\rho_{AB}\big]\ge0$, i.e., $\tr(H\rho_{AB})\le\tr(H_+\rho_{AB})$.
	By the Schatten-H\"{o}lder inequality, we have
	\begin{eqnarray}\label{lemma tr 6}
		\tr(H_+\rho_{AB})\le\|H_+\|_r\|\rho_{AB}\|_q.
	\end{eqnarray}
	Eqs.\eqref{lemma tr 4}-\eqref{lemma tr 6} yields
	\beax 
	\|\rho_A\|_q+\|\rho_B\|_q &\le 1+\|H_+\|_r\|\rho_{AB}\|_q
	< 1+\|\rho_{AB}\|_q. 
	\eeax
	Hence
	\beax
	1+\|\rho_{AB}\|_q = \|\rho_A\|_q+\|\rho_B\|_q \Rightarrow\rho_A\ \text{or}\ \rho_B \text{ is pure}.
	\eeax

	(ii) We need to check the other case of Eq.~\eqref{lemma tr}, i.e., $qs>1$.
	We let 
	\beax \left(\tr\rho_{AB}^{q}\right)^{1/q}=a, ~\left(\tr\rho_{A}^{q}\right)^{1/q}=b,~ \left(\tr\rho_{B}^{q}\right)^{1/q}=c.\eeax
	It is obvious that Eq.~\eqref{lemma tr} holds true for the cases of $a\ge b\ge c$, $a\ge c> b$, $b\ge a\ge c$, $c\ge a\ge b$, $b\ge c=a$, and $c> b=a$, respectively. So we only need to check the case when $b\ge c>a$ or $c> b>a$. We write $1+a=g$, $g=\lambda_1+\mu_1=\lambda_2+\mu_2$, where $\lambda_1\ne\lambda_2$ and $a<\lambda_1,\lambda_2 <\frac{g}{2}$. Considering 
	\beax f(\lambda)=\lambda^{p}+(g-\lambda)^p, ~\lambda\in (a, g/2), p>1,\eeax
	we have
	\begin{eqnarray}
		f'(\lambda)=p\lambda^{p-1}-p(g-\lambda)^{p-1}.\notag
	\end{eqnarray}
	So $f(\lambda)$ is strictly monotonically decreasing on $(a,\frac{g}{2})$. Therefore, when $\lambda_1>\lambda_2$,	$\lambda_1^{p}+\mu_1^{p}<\lambda_2^{p}+\mu_2^{p}$ holds true.
	Then for $\lambda_1>a$, $\lambda_1^{p}+\mu_1^{p}<1+a^{p}$.
	Since $\lambda_1+\mu_1=1+a\ge b+c$ and $a<c\le\frac{g}{2}$, if we take $\lambda_1=c$, then $\mu_1\ge b$, and
	\begin{eqnarray}
		1+a^p>\lambda_1^p+\mu_1^p\ge c^p+b^p.
	\end{eqnarray}
	From the argument above, if $1+a^p=b^p+c^p$, then either $b=1$ or $c=1$. This completes the proof since $p>1$ is equivalent to $qs>1$ in Eq.~\eqref{lemma tr}.

	\begin{widetext}

	\section*{Appendix 4: The proof of Proposition~\ref{Lower bipartiten}} \label{appendix-E}

	For any $n$-qudit pure state $\rho=|\psi\rangle\langle\psi|$, we have 
	\beax  \quad 
	&&\frac{1}{(q-1)s}\left[ 1-\left( \tr\rho_1^{q}\right) ^s\right] -\sum\limits_{\gamma_i \in \Gamma_2}E_{q,s}^{\gamma_i}(|\psi\rangle)-\dfrac{(2^{n-1}-2)(c_{d}-1)}{(q-1)s} \notag \\
	 \quad  &=&\frac{1}{(q-1)s}\left\lbrace \left[ 1-\left( \tr\rho_1^{q}\right)^s\right] -\sum\limits_{\gamma_i \in \Gamma_2}\left[ 1-\left( \tr\rho_{{X_{1(i)}}}^{q}\right)^s\right] -(2^{n-1}-2)(c_{d}-1)\right\rbrace \notag \\
	 \quad &=&-\frac{1}{(q-1)s}\left\lbrace \sum\limits_{\gamma_i\neq A_1|\overline{A_1}}\left[ 1-\left( \tr\rho_{{X_{1(i)}}}^{q}\right) ^s\right] +(2^{n-1}-2)(c_{d}-1)\right\rbrace
	\geq 0,
	\eeax
	where we have used the inequality $1-(\tr\rho_{{X_{1(i)}}}^{q})^s\le1-c_{d}$ for all partitions $\gamma_i\in\Gamma_2$ in the last inequality. Thus, for any bipartition $\gamma=X|Y\in\Gamma_2$, we get
	\be
	\frac{1}{(q-1)s}\left[ 1-(\tr\rho_{X}^{q})^s\right] \ge\sum\limits_{\gamma_i \in \Gamma_2}E_{q,s}^{\gamma_i}(|\psi\rangle)+\dfrac{(2^{n-1}-2)(c_{d}-1)}{(q-1)s}.
	\ee
	It follows from the definition of $E_{g(q,s)}^{(n)}$ that 
	\be
	E_{g(q,s)}^{(n)}(|\psi\rangle)\ge\sum\limits_{\gamma_i \in \Gamma_2}E_{q,s}^{\gamma_i}(|\psi\rangle)+\dfrac{(2^{n-1}-2)(c_{d}-1)}{(q-1)s}.
	\ee
	
	For the mixed $n$-qudit state $\rho$, we consider the optimal decomposition $\rho=\sum_{j}p_{j}|\psi_j\rangle\langle \psi_j|$ with respect to $E_{g(q,s)}^{(n)}$. We have
	\begin{align}
		E_{g(q,s)}^{(n)}(\rho)& =\sum_{\{p_j,|\psi_j\ra\}}p_{j}E_{g(q,s)}^{(n)}(|\psi_j\ra)
		\ge\sum_{j}p_{j}\left[ \sum\limits_{\gamma_i \in \Gamma_2}E_{q,s}^{\gamma_i}(|\psi_j\rangle)+\dfrac{(2^{n-1}-2)(c_{d}-1)}{(q-1)s}\right] \notag\\
		&\ge\sum\limits_{\gamma_i \in \Gamma_2}E_{s}^{\gamma_i}(\rho)+\dfrac{(2^{n-1}-2)(c_{d}-1)}{(q-1)s},\notag
	\end{align}
	which finishes the proof of Eq.~\eqref{lower1-1}. Eq.~\eqref{lower1-2} can be checked similarly.

	\section*{Appendix 5: The proof of Proposition~\ref{Lower multipartiten}} \label{appendix-F}

	We only need to prove Eq.~\eqref{lower2-1}, Eq.~\eqref{lower2-2} can be argued analogously. For any $n$-qudit pure state $\rho=|\psi\rangle\langle\psi|$, we have 
	\beax
	&&\frac{1}{(q-1)s}\left[ 1-(\tr\rho_1^{q})^s\right] -2E_{q,s}^{(n)}(\rho)-\dfrac{(n-1)(c_{d}-1)}{(q-1)s}\\
	&=&\frac{1}{(q-1)s}\left\lbrace 1-\left(\tr\rho_1^{q}\right)^s-\left[ n-\sum_{j=1}^{n}\left(\tr\rho_{j}^q\right)^s\right]-(n-1)(c_{d}-1)\right\rbrace \\
	&=&\frac{1}{(q-1)s}\left\lbrace -\left[ n-1-\sum_{j=2}^{n}\left( 1-\left( \tr\rho_j^{q}\right)^s\right) \right] -(n-1)(c_{d}-1)\right\rbrace \\
	&\ge&0,
	\eeax
	since $1-(\tr\rho_j^{q})^s\le1-c_{d}$ for $j=2,3,\cdots,n$. Thus, we get
	\begin{eqnarray}
		\frac{1}{(q-1)s}\left[ 1-\left( \tr\rho_j^{q}\right)^s\right] \ge2E_{q,s}^{(n)}(|\psi\rangle)+\dfrac{(n-1)(c_{d}-1)}{(q-1)s},
	\end{eqnarray}
	for $j=2,3,\cdots,n$.
	This implies
	\begin{eqnarray}
		E_{g(q,s)}^{(n)}(|\psi\rangle)\ge2E_{q,s}^{(n)}(\rho)+\dfrac{(n-1)(c_{d}-1)}{(q-1)s}.
	\end{eqnarray}
	
	For the mixed $n$-qudit state $\rho$,
	\beax
	E_{g(q,s)}^{(n)}(\rho)=\sum_{\{p_j,|\psi_j\ra\}}p_{j}E_{g(q,s)}^{(n)}(|\psi_j\ra)\ge\sum_{j}p_{j}\left[ 2E_{q,s}^{(n)}(|\psi_j\rangle)+\dfrac{(n-1)(c_{d}-1)}{(q-1)s}\right]\ge2E_{q,s}^{(n)}(\rho)+\dfrac{(n-1)(c_{d}-1)}{(q-1)s},
	\eeax
	which completes the proof. 	
	
\end{widetext}


	
\end{document}